\documentclass[singlecolumn,aps,prl]{revtex4-1}

\usepackage{amsmath}
\usepackage{graphicx}
\usepackage{esint}
\usepackage{epsfig}

\usepackage{color}

\def\beq{\begin{equation}}
\def\eeq{\end{equation}}
\def\beqa{\begin{eqnarray}}
\def\eeqa{\end{eqnarray}}
\def\e{\epsilon}

\def\e{\epsilon}
\def\cH{{\mathcal H}}

\def\mx{\mathrm{mx}}
\def\d{\mathrm{d}}
\def\rmL{\mathrm{L}}
\def\rmR{\mathrm{R}}

\def\nonum{\nonumber \\}

\def\nonum{ \nonumber \\}
\def\emx{\eta_{\mathrm{mx}}}
\renewcommand{\sec}[1]{\vskip 0.5truecm \noindent \emph{#1}. -- }

\begin{document}  

%
\author{Shigeru Ajisaka}
\affiliation{Department of Chemistry, Ben-Gurion University of the Negev, Beer-Sheva 84105, Israel}
\author{Bojan \v Zunkovi\v c}
\affiliation{Departamento de F\' isica, Facultad de Ciencias F\' isicas y Matem\' aticas, Universidad de Chile, Casilla 487-3, Santiago Chile}

\author{Yonatan Dubi $^*$} 
\affiliation{Department of Chemistry and Ilse-Katz Institute for Nanoscale Science and Technology, Ben-Gurion University of the Negev, Beer-Sheva 84105, Israel}

\title{The Molecular Photo-Cell: Quantum Transport and Energy Conversion at Strong Non-Equilibrium}
\date{\today}


\begin{abstract}
The molecular photo-cell is a single molecular donor-acceptor complex attached to electrodes and subject to external illumination. Besides the obvious relevance to 
molecular photo-voltaics, the molecular photo-cell is of interest 
being a paradigmatic example for a system that inherently operates in out-of-equilibrium conditions and typically far from the linear response regime. Moreover, this system 
includes electrons, phonons and photons, and environments which 
induce coherent and incoherent processes, making it a challenging system to address theoretically.  Here, using an open quantum systems approach, we analyze the 
non-equilibrium transport properties and energy conversion performance of a molecular photo-cell, including for the first time both coherent and incoherent 
processes and treating electrons, photons, and phonons on an equal footing. We find that both the non-equilibrium conditions and decoherence play a crucial role in 
determining the performance of the photovoltaic conversion and the optimal energy configuration of the molecular system. 
\\
\\ \\
$^*$ Correspondence to jdubi@bgu.ac.il 

\end{abstract}
\maketitle   


\section{Introduction} 
Understanding the properties of non-equilibrium systems has been a central effort of the scientific community for many years. Of specific interest are non-equilibrium processes that 
take place at the nano-meter scale and at which energy is converted from one form to another, for instance photovoltaic (PV) energy 
conversion, photochemistry and photosynthesis. In these cases, the interaction between electrons and photons under non-equilibrium conditions plays an essential role. Theoretical 
modeling of such processes is a challenging task, since the interacting nature of the system and its many-body characteristics, the multitude of constituents, the presence of external 
environments, and the non-equilibrium conditions must all be taken into consideration. 

Even harder to address theoretically are situations in which the system has two independent fluxes, originating from separate non-equilibrium drivings, and are both far away from the 
linear response regime, a situation which is designated as {\sl strong  
non-equilibrium}. A paradigmatic example for such a system are photo-voltaic cells, where the two fluxes are the heat flux, originating from the huge temperature difference 
between the sun and the earth, and the particle current originating from the voltage difference between the electrodes. In recent years a new and exciting class of photo-voltaic cells has emerged, namely molecular photo-cells, where the energy conversion process takes place at the single molecule level \cite{Gratzel2005,Deibel2010,Nicholson2010}. 

Here we propose a formalism to study non-equilibrium transport in molecular junctions, and use it to investigate a model for the molecular photo-cell, a single molecular donor-acceptor complex attached to electrodes and subject to external illumination. This model was recently 
suggested \cite{Einax2011,Einax2013,Smirnov2009} to be the minimal model to describe PV energy conversion in ideal, single-molecule heterojunction organic PV cells. In Ref.~\cite{Einax2011,Einax2013} PV conversion efficiency was analyzed using the (essentially classical) rate equations for the 
electronic degrees of freedom. The dynamics and non-equilibrium properties of the phonons and photons were ignored, being considered only within a (non-self-consistent) mean-field approximation and assumed to have equilibrium 
distributions. Here we show the non-equilibrium properties of the phonons and photons have a strong impact on the PV conversion properties in realistic parameter range and cannot be neglected. 

The formalism we present here allows us to treat electrons, photons and phonons fully 
quantum mechanically and on an equal footing (without resorting to a mean-field approximation) and to take into account the action of 
the environments producing a strong non-equilibrium situation. We use the many-body Lindblad quantum master equation \cite{Kampen2007,Lindblad1976,Gorini1976} to describe the environments, which consist of metallic electrodes in touch with a molecular complex, a phonon bath (at ambient temperature) and a photon bath (at the solar temperature). Non-equilibrium is induced by two sources, namely the temperature difference between the incoming photons (originating from the sun) and the device, and the voltage bias between the electrodes. 

Using this formalism, we calculate the non-equilibrium  densities of electrons, photons and phonons, the electric current and power output, and the 
thermodynamic efficiency at maximal power. We find that under certain conditions, the distribution functions for the phonons and photons can be very different from the equilibrium distributions, 
and therefore approximating the system as close to equilibrium is not a valid approximation. We then study the signature of non-equilibrium on the energy conversion efficiency of the system. 

In addition to be able to include strong non-equilibrium effects and to account for all constituents on equal footing, our formalism allows one to introduce effects of dephasing and decoherence in a simple and physically transparent way. We study how the loss of quantum coherence affects the efficiency, and show that classical 
electron transfer (as opposed to coherent electron propagation) {\sl enhances} the efficiency.  

\section{Results: efficiency for the coherent system far from equilibrium} 

\begin{figure}[h!]
\vskip 0.5truecm
\includegraphics[width=6truecm]{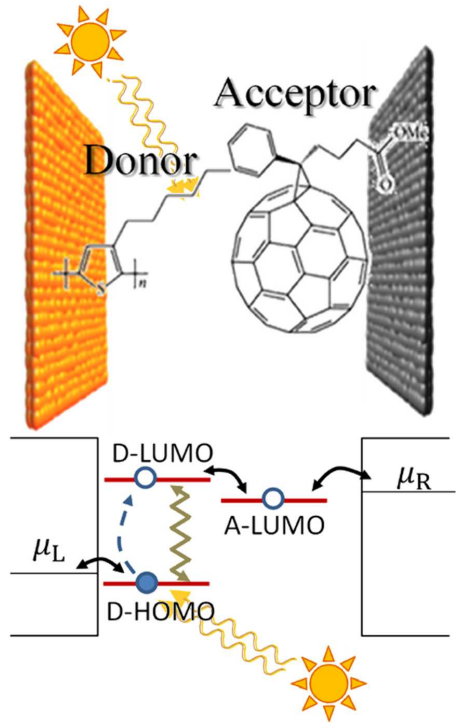}
\caption{Schematic illustration of the minimal model for a molecular PV cell. The system consists of a molecule donor and an acceptor molecule, characterized by their HOMO and LUMO levels and coupled to each other via electron hopping. The D-molecule is coupled only to the left electrode, and the A-molecule only to the right electrode. Electrons in the donor interact with both photons (wiggly line) and phonons (broken line).}
\label{fig1}
\end{figure}

The system under consideration is composed of a molecular junction, in which a donor-acceptor (D-A)
complex is placed between two metallic electrodes (Fig.~\ref{fig1}) \cite{Einax2011}. This is an idealization of an envisioned future single-layer molecular photovoltaic cells, where a self-assembled layer of D-A pairs is placed on a conducting substrate, and is covered by a top transparent electrode. For the donor (D), we consider the highest occupied molecular orbital (HOMO) with energy $\epsilon_{D,1}$ and lowest 
unoccupied molecular orbital (LUMO) with energy $\epsilon_{D,2}$. For the acceptor (A), we consider only the LUMO with energy $\epsilon_{A}$ (the energy of the A's HOMO is much 
lower and is unaffected by any dynamics in the photocell \cite{Einax2011}). Recent advances in the experimental ability to measure photo-conductivity and PV conversion in single-molecule junctions \cite{battacharyya2011,fereiro2013,furmansky2012,aradhya2013,gerster2012} make our theoretical model experimentally relevant.

\begin{figure}[h!]
\vskip 0.0truecm
\includegraphics[width=7.5truecm]{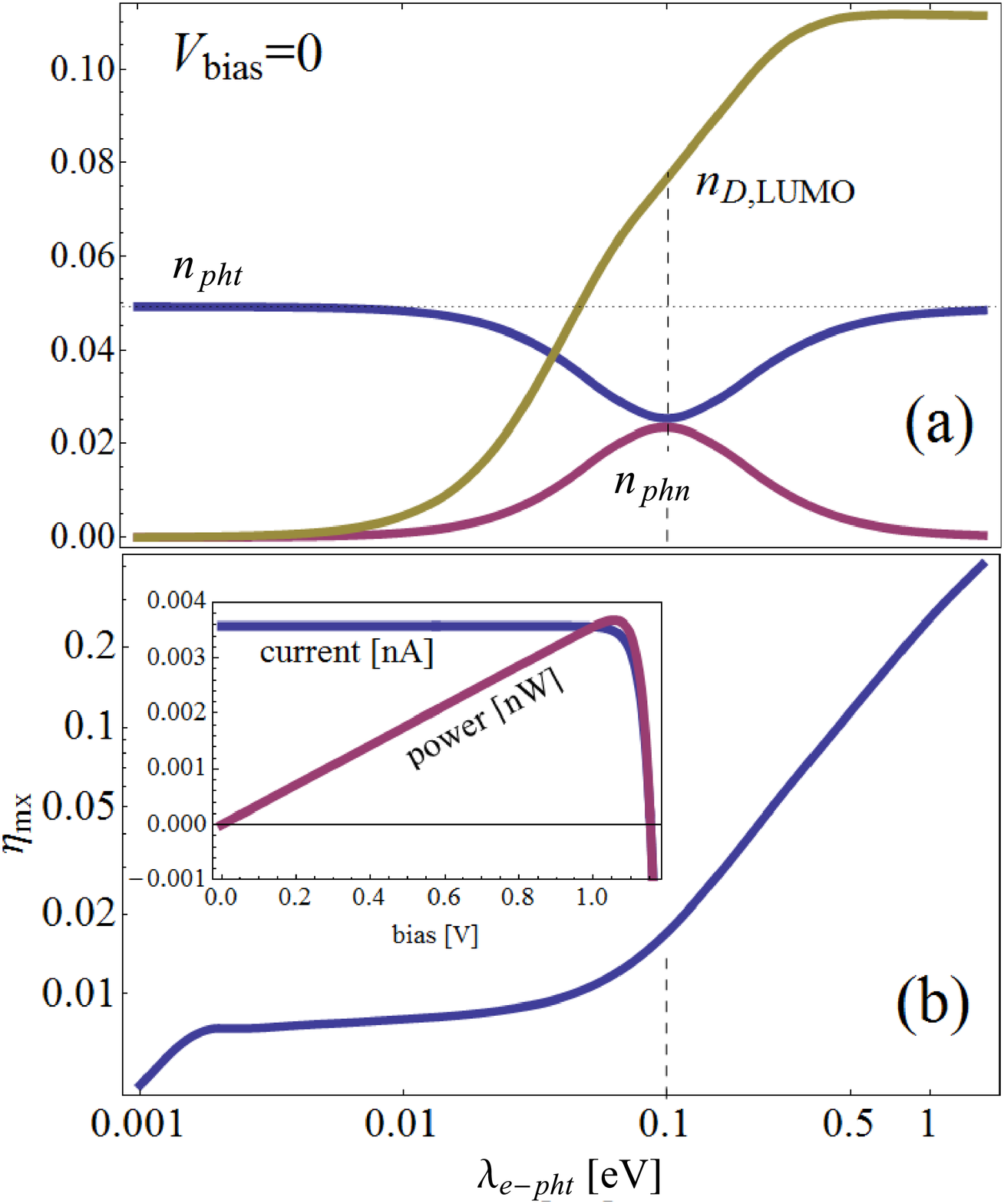}
\caption{(a) Photon density $n_{\mathrm{pht}}$, phonon density $n_{\mathrm{phn}}$ and the electronic 
occupation of the donor LUMO as a function of $e-pht$ coupling $\lambda_{e-pht}$. Dashed line indicates the value of $\lambda_{e-phn}$. Dotted horizontal line marks the equilibrium occupation of solar photons.(b) Efficiency at maximum power $\emx$ as a function of the $e-pht$  coupling (same parameters as in (a)). Inset: typical current $J$ and power output $P=J\times V$ vs. bias voltage $V$. }
\label{fig2}
\end{figure}

We begin by examining the electron, photon and phonon densities at zero bias voltage. Note that the system is still out of equilibrium  
due to the temperature difference between the photon and phonon baths. In Fig.~\ref{fig2}(a) the photon density $n_{\mathrm{pht}}$, the phonon density $n_{\mathrm{phn}}$, and the electronic 
occupation of the D-LUMO, $n_{\mathrm{D,LUMO}}$, are plotted as a function of $e-pht$ coupling $\lambda_{e-pht}$. We set the orbital energies to be $\epsilon_{D,1}=-0.1$ eV, $\epsilon_{D,2}=1.4$eV, and $\e_A=1.25$eV 
\cite{Einax2011}. The $e-phn$ interaction is fixed at $\lambda_{e-phn}=0.1$ eV (dashed vertical line in Fig.~\ref{fig2}(a)). Three distinct regimes are observed: (i) At small $e-pht$ coupling, 
 $\lambda_{e-pht} \ll  \lambda_{e-phn}  $, the system is close to equilibrium, and the photon occupation is defined by the solar temperature $T_s$ (dotted line). The phonons are excited 
according to the ambient temperature (which is very small compared to $\omega_0$, and consequently the phonon occupation is very small), and the occupation of the D-LUMO level is also very small. (ii) As the $e-pht$ and $e-phn$ interactions 
become comparable, energy is transferred from the photons to the phonons, mediated by excitation of electrons from the D-HOMO to the D-LUMO level. As a result, the D-LUMO and phonon occupations 
increase, while the photon occupation decreases. Alternatively, this situation can be described in terms of heating (although the notion of temperature is not applicable out of equilibrium, it is still useful to think in terms of an effective temperature): the junction is locally (and efficiently) heated by the photons. This heat is transferred to the phonons, resulting in an elevated effective phonon temperature and, consequently, enhanced phonon occupation. (iii) When $e-pht$ coupling is large,  $\lambda_{e-pht} \gg \lambda_{e-phn}$, there is no longer efficient transfer of 
heat to the phonons. However, the D-LUMO occupation continues to rise, due to energy pumping from
the photons to the D-LUMO. 

An important measure of the operational efficiency of the molecular PV cell is the efficiency at maximal power, $\emx$, defined as the ratio between the cell's maximal power output $P_{\mathrm{out,mx}}$ and the corresponding input power $P_{\mathrm{in}}$ supplied by 
the photons  \cite{Esposito2009a,VandenBroeck2005}. The output power is given by $P_{\mathrm{out}}=J \times V$, where $J$ is the particle current through the system, and the input power is calculated in a 
similar way \cite{Einax2011} from $P_{\mathrm{in}}=2i\omega_0 \lambda_{e-pht}  \langle a^{\dagger} c^\dagger_{D,1} c_{D,2} -a c^\dagger_{D,2} c_{D,1} \rangle $, which is related to the 
photon-induced part of the particle current.

As shown in the inset of Fig.~\ref{fig2}(b), a typical $J-V$ curve has a maximal power.
In Fig.~\ref{fig2}(b), $\emx$ at the maximal power is plotted as a function of the $e-pht$  coupling (we use the same parameters as in Fig.~\ref{fig2}(a)). For very small $e-pht$ coupling ($\lambda_{e-pht}<$0.002 eV), 
the efficiency is very small, and grows linearly with $\lambda_{e-pht}$. In this regime, the time it takes for an electron to absorb a photon is larger than the time the photons spent in the cell (defined by 
$\gamma_{pht}n_B(T_S)$ (which translates to $\sim 0.002$ eV), and so the photon absorption is very small leading to poor efficiency. This regime is followed by a plateau regime, where any energy transfered from the 
photons to the electrons is quickly dissipated by phonons and is not converted into electrical power. There is thus little change in the efficiency, as long as the rate of photon absorption is smaller than the 
eletron-photon-relaxation time. Only when the $e-pht$ coupling reaches the $e-phn$ coupling $\lambda_{e-phn}=0.1$ eV the efficiency begins to increase: in this regime, energy is transferred to the electrons by the photons 
faster than can be dissipated by the phonons, and as a result an increasing amount of this energy is transferred into electronic power, resulting in a rise of efficiency. 

\begin{figure}[h!]
\vskip 0.0truecm
\includegraphics[width=6.5truecm]{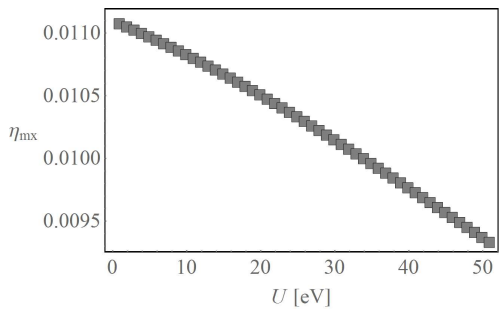}
\caption{Efficiency at maximum power $\emx$ as a function of the exciton Coulomb energy $U$ }
\label{fig2_c}
\end{figure}

It is important to note that the results described in Fig.~\ref{fig2}, especially in the region where the electron-phonon and electron-photon couplings are of the same order, {\sl cannot be obtained by assuming equilibrium distributions for the phonons and photons}, and this situation is described here for the first time. Since in future realistic devices the strength of 
the electron-photon and electron-phonon interactions are unknown, a situation where they are of similar magnitude may occur, in which case the system dynamics cannot be described as close to equilibrium, and the full non-equilibrium dynamics need to be taken into account. 
To further demonstrate the power of this method, we next discuss the effect of Coulomb interactions on the efficiency. In the Hamiltonian of Eq.~\ref{Hamiltonian}, the A-LUMO energy, $\e_A$, 
already includes the Coulomb repulsion energy on the acceptor \cite{Einax2011}. In Excitonic systems, the Coulomb interaction is typically considered through the "exciton binding energy", defined by the Coulomb 
interaction term $\cH_{C}=U c^{\dagger}_{D,2}c_{D,2}\left (1-c^{\dagger}_{D,1}c_{D,1}\right)$ between an electron at the D-LUMO and a hole in the D-HOMO. While in methods such as non-equilibrium 
Green's function adding Coulomb interaction requires substantial effort, the present method does not require either additional technical complexity or additional 
computational power to account for any Coulomb interaction effects. In Fig.~\ref{fig2_c} we show the efficiency at maximum power $\eta$ as a function of the exciton Coulomb energy $U$ (which can be estimated from, e.g. density-functional calculations). We set $\lambda_{e-pht}=0.1$eV and $\lambda_{e-phn}=0.2$eV. We find an almost linear decrease in the efficiency, with a reduction of $\sim 15 \% $ for $U=0.2$eV.

\section{Results: the role of decoherence} 

The next question we wish to address is the extent to which the quantum nature of the system affects the PV conversion efficiency, a question which is beyond the reach 
of the formalism presented in Refs.~\cite{Einax2011,Einax2013}. 
The formalism we present here allows us to access, in addition to fully quantum-coherent processes described above, also incoherent processes. The most important incoherent processes are electron 
transfer from the D-LUMO to the A-LUMO, described classically in Ref. \cite{Einax2011}. These are addressed here by adding an additional pair of $\hat{V}-$operators that accounts for incoherent 
transitions, namely $\hat{V}_{D\rightarrow A}= \Gamma^{1/2}_{D-A}  c^\dagger_{D,2} c_A,~~\hat{V}_{A\rightarrow 
D}= \Gamma^{1/2}_{A-D} c^\dagger_{A} c_{D,2}$, where $\Gamma_{D-A}=\gamma_{D-A}\times \left(1+\exp\left(\frac{\e_{A}-\e_{D,2}}{k_B T}  \right)  \right)^{-1}$. Thus, the pair of parameters 
$t_{D-A}$ and $\gamma_{D-A}$ describe the strength of the coherent and incoherent donor-acceptor electron transfer processes, respectively.

In what was shown in Fig.~\ref{fig2}, the D- and A- LUMO levels were connected by quantum-mechanical bonding. In 
contrast, Ref.~\cite{Einax2011} accounted for the electronic transfer between the D- and A- LUMO levels by an incoherent (or classical) transfer process. It is thus of 
interest to interpolate between the fully quantum case ($t_{D-A}\neq 0,\gamma_{D-A}=0$), through the mixed quantum-classical case ($t_{D-A}\neq 0,\gamma_{D-A}\neq 0$), to the fully classical case ($t_{D-A}=0 ,\gamma_{D-A}\neq 0$). 

To do so, we define a variable $\xi$ such that $0\leq \xi \leq 1$, and define $t_\mathrm{mx}=0.05$ eV and $\gamma_\mathrm{mx}=10^{12} s^{-1}$ (as in Ref.~\cite{Einax2011}). We now parametrize $t_{D-A}$ and $\gamma_{D-A}$ with $\xi$
, $t_{D-A}=t_\mathrm{mx} \left(1-2(\xi-0.5)\Theta(\xi-0.5)  \right), ~ \gamma_{D-A}=\gamma_\mathrm{mx} \left(1-2(0.5-\xi)\Theta(0.5-\xi)  \right)$  ($\Theta(\xi)$ is the Heaviside unit step-function). This parametrization is 
shown on the right inset of Fig.~\ref{fig3}, and is constructed such that for $\xi=0$ the system is fully coherent, for $\xi=0.5$ the system is mixed (both quantum and classical processes), and for $\xi=1$ the system is fully 
incoherent, so the range $0<\xi<1$ interpolates between all three cases.

In Fig. ~\ref{fig3} we plot the efficiency at maximum power 
$\emx$ as a function of the position of the A-LUMO, $\epsilon_A$ and the parameter $\xi$. We set $\gamma_{D-A}=10^{12} s^{-1}$ as in Ref.~\cite{Einax2011}. We find that the quantum coherence 
or classical decoherence (parametrized by $\xi$) has a profound effect on the efficiency of the molecular PV-cell in two important aspects. 

First, the optimal position of the A-LUMO energy differs according to the nature of the transition under consideration: quantum (coherent), both quantum and classical, or classical D-A transitions (solid lines in Fig. ~\ref{fig3}). For the last case  ($t_{D-A}=0 ,\gamma_{D-A}= 10^{12} \mathrm{s^{-1}}$), we find that  $\e_A$ is optimal at $\e_A \sim 1.25$ eV, verifying the result of Ref.~\cite{Einax2011}. When quantum correlations are added (
$t_{D-A}=0.05 \mathrm{ eV} ,\gamma_{D-A}=10^{12} \mathrm{s^{-1}}$), two peaks emerge at $\e_A \sim 0.9, 1.3$ eV, and a lower peak emerges at $\e_A \sim 1.6$ eV. For a system with only 
quantum transitions ($t_{D-A}=0.05 \mathrm{ eV} ,\gamma_{D-A}= 0$, enlarged in the back inset in Fig.~\ref{fig3}), the lower peak vanishes, and the optimal LUMO positions are at $1.2$ eV and $1.6$ eV. Thus, in the design of optimal molecular PV cells, it is important to take into account the quantum nature. 

Second, as can be clearly seen in Fig.~\ref{fig3}, the addition of classical D-A transitions {\sl increases} the efficiency substantially by more than an order of magnitude. This finding is surprising, since one would 
expect that incoherent (and dissipative) transitions would lead 
to a decrease in the efficiency. To understand the origin of this effect, we performed time-dependent calculations (not shown) and found that for the coherent case, an electron that is 
excited to the D-LUMO coherently oscillates between the D- and A-LUMO, while for the incoherent case the electron decays from the D- to the A-LUMO exponentially (and its return rate is exponentially small). This implies that in the coherent case the electron spends much more time in the D-LUMO than in the incoherent case, before transferring to the right electrode.
Since the electron can decay back to the D-HOMO (emitting a phonon) only directly from the D-LUMO, the longer it spends on the D-LUMO, the higher the probability for non-radiative decay back 
to the D-HOMO, leading to a decrease in efficiency. This phenomena is similar to dephasing-assisted transport conjectured to occur in biological systems,  \cite
{Caruso2009,Mohseni2008,Rebentrost2009,Plenio2008}  but here is the first time it is discussed and demonstrated in the context of molecular PV cells. Since in realistic single-molecule both coherent and incoherent effects may be important, they must be included in a theoretical description of the system. 

\begin{widetext}
\begin{center}
\begin{figure}[h!]
\vskip 0.0truecm
\includegraphics[width=12.5truecm]{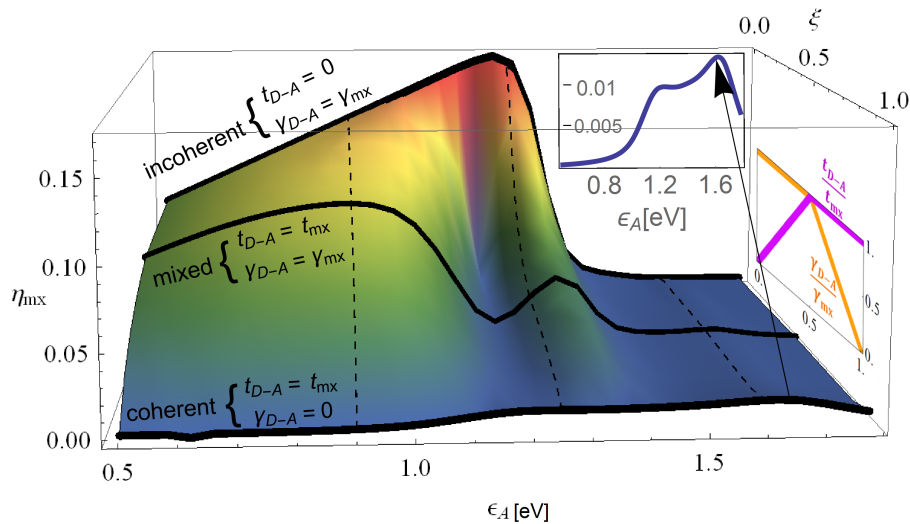}
\caption{Efficiency at maximum power $\emx$ as a function of the position of the A-LUMO $\epsilon_A$ and the parameter $\xi$ which describes the interpolation from 
coherent to incoherent electron D-A processes (see text). Solid lines mark the coherent ($t_{D-A}=t_\mx,\gamma_{D-A}=0$), mixed ($t_{D-A}=t_\mx,\gamma_{D-A}=\gamma_\mx $) 
and incoherent ($t_{D-A}=0,\gamma_{D-A}=\gamma_\mx$) donor-acceptor electron transfer processes. Dashed lines are guides to the eye, showing the position of the maximal 
efficiency for the different cases. 
Right inset: parametrization of $t_{D-A}$ and $\gamma_{D-A}$ with $\xi$. Back inset: Efficiency as a function of the position of the A-LUMO $\epsilon_A$ at $\xi=0$ (fully coherent system).}
\label{fig3}
\end{figure}

\end{center}
\end{widetext}
To illustrate this connection between dynamics and efficiency, we address the relaxation dynamics of the donor-acceptor system. Considering only the D-A LUMOs and the right electrode (without the photons and phonons), we construct the Lindbladian $\hat{M}$-matrix of Eq.~\ref{Lindbladian} (at zero bias), by constructing a vector form for the Lindblad equation of Eq.~\ref{Lindbladian}, $\dot{\vec{\rho}} =-\hat{M} \vec{\rho}$. The $\hat{M}$-matrix has a zero eigenvalue, which defines the steady state. The (real part of the) rest of the eigenvalues define the relaxation rates towards the steady state. The minimal rate $\Gamma_{\mathrm{min}}$ (i.e., eigenvalues of $\hat{M}$ with smallest real part which, we numerically check, is non-zero) represents 
the longest relaxation time for the system to reach the steady state from any general state. 

\begin{figure}[h!]
\vskip 0.0truecm
\includegraphics[width=6.5truecm]{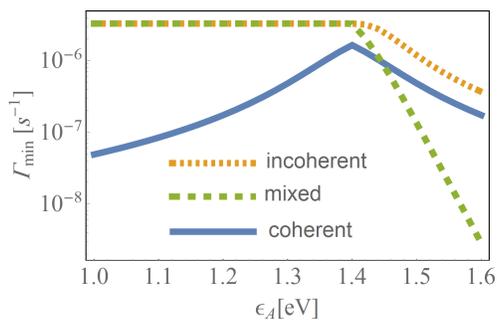}
\caption{ Minimal decay rate $\Gamma_{min}$ (on a log scale) as a function of the position of the acceptor LUMO, $\epsilon_A$, for the three cases of fully coherent ($t_{D-A}=t_\mx,\gamma_{D-A}=0$), mixed ($t_{D-A}=t_\mx,\gamma_{D-A}=\gamma_\mx$) 
and incoherent ($t_{D-A}=0,\gamma_{D-A}=\gamma_\mx$) donor-acceptor electron transfer processes.}
\label{fig5}
\end{figure}

 In Fig.~\ref{fig5}, we plot the decay rate $\Gamma_{\mathrm{min}}$ of the model which only contains D-A LUMOs and the right electrode (without the photons and phonons) as a function of the position of the A-LUMO $\epsilon_A$ for the three cases of fully coherent ($t_{D-A}=t_\mx,\gamma_{D-A}=0$), mixed ($t_{D-A}=t_\mx,\gamma_{D-A}=\gamma_\mx$) 
and incoherent ($t_{D-A}=0,\gamma_{D-A}=\gamma_\mx$) donor-acceptor electron transfer processes. As can be 
seen, the decay rate is much smaller for the fully coherent case, indicating a longer relaxation time. This is in line with the observation above that slower relaxation dynamics lead to lower efficiency. 
In addition, we point out that the relaxation times (which are much simpler to calculate than the full efficiency) serve as an indicator for the efficiency, even though they do not capture the fine details required for an optimal design of the system (see Fig.~\ref{fig3}(a) and (b)). 

\begin{figure}[h!]
\vskip -0.5truecm
\includegraphics[width=8.5truecm]{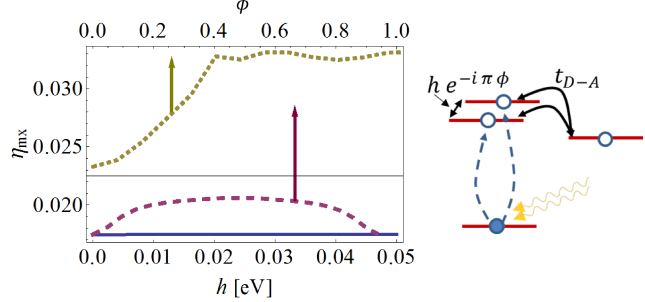}
\caption{Right: Schematic illustration of the molecular PV cell, with two donor LUMO levels. Left: Efficiency at maximum power $\emx$ as a function of hopping amplitude $h$ between the donor LUMO levels and the acceptor LUMO (solid line), and hopping matrix element phase $\phi$ (dashed line), indicating the effect of quantum coherence on the efficiency for fully coherent D-A transfer ($\gamma_{D-A}=0$). The dotted line is the same for a mixed coherent-incoherent transfer ($\gamma_{D-A}\neq 0$).
}
\label{fig6}
\end{figure}

To further examine possible effects of coherence on the efficiency of the molecular PV cell, we study a system where the donor has two degenerate D-LUMO levels which have been introduced experimentally \cite{Rizzi2008,Terazono2007}. Here, we study a simplified system (schematically depicted on the right side of Fig.~\ref{fig6}), in which the photons and the phonons excite electrons with equal amplitudes from the D-HOMO to the two D-LUMO 
levels. Each of the levels is coupled to the A-LUMO with the same hopping amplitude $t$, and they are coupled to each other with a complex hopping amplitude $h e^{-i \pi 
\phi}$. In 
Fig.~\ref{fig6}, the efficiency at maximum power $\emx$ is plotted as a function of the inter-LUMO coupling $h$ (solid line) and the phase $\phi$ (dashed line) for 
$\gamma_{D-A}=0$, i.e., no incoherent D-A transfer. We find that while $h$ has little effect on the efficiency, the phase $\phi$ has a significant effect (increasing the 
efficiency by up to $\sim 15\%$). Surprisingly, we also find that this quantum interference effect persists even when incoherent D-A transfer is included 
($\gamma_{D-A}=10^{12} s^{-1}$ as in Fig.~\ref{fig3}), and an substantial increase of $\eta_{\mx}\sim 30\%$ is observed by varying $\phi$ (dotted line).

\sec{Discussion} In Summary, we have proposed a novel formalism to study non-equilibrium quantum transport in molecular junctions, and applied it to investigate a minimal model of PV energy 
conversion in ideal, single-molecule PV cells. The results shown above indicate that quantum coherence effects are important in 
determining the non-equilibrium energy conversion performance of molecular PV cells. The formalism presented here sets the stage for a fully coherent quantum mechanical calculation of energy 
conversion in more realistic models for molecular PV cells and can be directly linked to quantum chemistry methods (such as density-functional theory). The progress 
in the experimental ability to measure photo-conductivity and PV conversion in single-molecule junctions \cite{battacharyya2011,fereiro2013,furmansky2012,aradhya2013,gerster2012} allows one 
to envision real PV devices composed of a single molecular junction or a molecular monolayer, making our theoretical model experimentally relevant. Furthermore, our method can include both 
coherent 
and incoherent effects, making it a useful tool in the study of other energy conversion processes such as photosynthesis, where both classical and quantum processes take place \cite{Abramavicius2008,Ai2012,Caruso2009,Cheng2009,Chin2012,Collini2010,Engel2007a,Hoyer2010,Ishizaki2012,Lambert2013,Lee2007,Mohseni2008,Olaya-Castro2008,Plenio2008,Read2008,Rebentrost2009,Scholes2007,Calhoun2009,Caruso2010}, or other chemical and photo-chemical processes \cite{Miller2012,Xie2012}.

\section{Methods}

The full Hamiltonian of the molecular PV cell, including the molecular orbitals, the photons and the phonons, may be written as   
$ \cH =\cH_M+\cH_{pht}+\cH_{phn}+\cH_{M-pht}+\cH_{M-phn} $ where
$\cH_M$ is the Hamiltonian for the molecular complex, $\cH_{pht(phn)}$ is the photon (phonon) Hamiltonian, and $\cH_{M-pht(phn)}$ describes the electron-photon (phonon) interaction (we set $\hbar=1$ hereafter),
\beqa 
\cH_M &=& \sum_{x}\e_{x} c^{\dagger}_{x}c_{x}-t_{D-A}(c^{\dagger}_{D,2}c_A+{\mathrm H.c}), \nonum 
\cH_{pht} &=& \omega_0 a^\dagger a,~~~\cH_{phn}= \omega_0 b^\dagger b ,\nonum
\cH_{M-pht} &=& \lambda_{e-pht} a^\dagger c^\dagger_{D,1}c_{D,2}+{\mathrm H.c.} ,\nonum
\cH_{M-phn} &=& \lambda_{e-phn} b^\dagger c^\dagger_{D,1}c_{D,2}+{\mathrm H.c.}~~.
\label{Hamiltonian}\eeqa
Here $c^\dagger_x (c_x)$ creates (annihilates) an electron in the D-HOMO ($x=D,1$), D-LUMO state ($x=D,2$) or A-LUMO state ($x=A$), with the 
corresponding level energies $\epsilon_x$, $a^\dagger (a)$ creates (annihilates) a photon with energy $\omega_0 =\epsilon_{D,2}-\epsilon_{D,1}$, and  $b^\dagger (b)$ 
creates (annihilates) a phonon with the same energy. In principle one should consider many photon (and phonon) modes, however the strongest effect on the dynamics comes from the resonant photons (with energy same as the HOMO-LUMO gap). The electron-photon Hamiltonian $\cH_{M-pht}$ describes (within the rotating wave approximation) the process (and its reverse process) in which an electron in the 
D-LUMO state relaxes to the D-HOMO state and emits a photon,  with the electron-photon ($e-pht$) coupling $\lambda_{e-pht}$. The electron-phonon Hamiltonian $\cH_{M-phn}$ is similar to $\cH_{M-pht}$,but with phonons 
instead of photons. For simplicity we consider spinless electrons, since  spin does not play a significant role in the energy conversion process. 


To study the dynamics of the system, we use the Lindblad equation to model the system and the environments \cite{Gorini1976,Lindblad1976,Breuer2002,DiVentra2008,Kampen2007}. The essence of the Lindblad approach is that instead of describing the environment by encoding it into a self-energy (as is done in the non-equilibrium Green's function approach \cite{DiVentra2008}) 
the environment is characterized by its action on the system. This action is mapped onto so-called 
Lindblad $\hat{V}$-operators, which describe incoherent transitions of the system elements due to the presence of an environment. The Lindblad equation was recently employed to address various aspects of 
electron transport \cite{Ajisaka2013,Ajisaka2012,Dzhioev2011a,Dzhioev2014a,dzhioev2011second,Dzhioev2011,Harbola2006}, yet in these studies the interaction with an environment was limited to electrons only, and the non-equilibrium dynamics of other constituents (i.e. phonons or photons) was not considered.

We assume that the left electrode is coupled only to the D-HOMO and that the 
right electrode is coupled only to the A-LUMO \cite{Einax2011}, as in Fig.~\ref{fig1}. The corresponding $\hat{V}$-operators are then \cite{Harbola2006,Ajisaka2012,Dzhioev2011}: 
\beqa
\hat{V}_{\rmL,+}&=&\sqrt{\gamma_L f_L(\epsilon_{D,1})} c^{\dagger}_{D,1},~
\hat{V}_{\rmL,-}=\sqrt{\gamma_L \tilde{f}_L(\epsilon_{D,1})} c_{D,1}~,\nonum
\hat{V}_{\rmR,+}&=&\sqrt{\gamma_R f_R(\epsilon_{A})} c^{\dagger}_{A},~
\hat{V}_{\rmR,-}=\sqrt{\gamma_R \tilde{f}_R(\epsilon_{A})} c_{A}~,
\label{Vops1} 
\eeqa
where $\gamma_{L,R}$ are electron transfer rates to the left and right electrodes,  
$T$ is the ambient temperature (we take $T=300$ K), $\mu_{L}=0$ is the left-electrode chemical potential, $\mu_R=V$ is the right electrode chemical potential, $V$ is the voltage bias, $f_{L,R}(\epsilon)=\left(1+\exp\left(\frac{\epsilon-\mu_{L,R}}{T} \right) \right)^{-1}$ are the Fermi-Dirac distributions of the left and right electrodes, and $\tilde{f}=1-f$. 
Following Ref.~\cite{Einax2011,Einax2013} we set $\gamma_L=\gamma_R=10^{10} s^{-1}$ (which corresponds to an energy scale of $\sim 4\times 10^-5$ eV, much smaller than the other electronic energy scales, validating the 
use of the Lindblad equations). 

For the bosons (photons and phonons), similar $\hat{V}$-operators that relate to the Bose-Einstein statistics of the boson baths are constructed, 
\beqa
\hat{V}_{pht,+}&=& \sqrt{\gamma_{pht} n_B(T_s)} a^\dagger ,~
\hat{V}_{pht,-}= \sqrt{\gamma_{pht} \tilde{n}_B(T_s)} a ~,\nonum
\hat{V}_{phn,+}&=& \sqrt{\gamma_{phn} n_B(T)} b^\dagger,~
\hat{V}_{pht,+}= \sqrt{\gamma_{phn}\tilde{n}_B(T)} b ~,\nonum
\label{Vops2}\eeqa
where $\gamma_{pht},\gamma_{phn}$ are photon and phonon relaxation rates (set to $\gamma_{pht}=\gamma_{phn}=10^{12} s^{-1}$), $T_s\sim 5700$ K is the solar temperature,  $n_B(\e)=\left(1-\exp\left(\frac{\epsilon}{T} \right) \right)^{-1}$ is the Bose-Einstein distribution, and $\tilde{n}_B=1+n_B$.


Once the $\hat{V}$-operators are set, the dynamics are determined by the propagation of the density matrix via the Lindblad equation, 
\beqa 
\dot{\rho}&=&-\frac{i}{h}[\cH,\rho ]+\sum_{ \hat{V} } \left(  -\frac{1}{2}\left\{ \hat{V}^\dagger\hat{V},\rho \right\}   +\hat{V}\rho\hat{V}^\dagger \right) ~~, \label{Lindbladian}
\eeqa
where $[ \cdot,\cdot]$ is the commutator and $\{ \cdot,\cdot \}$ is the anti-commutator. 
We numerically study the steady-state density matrix $\rho_{SS}$ with truncated bosonic space containing $n$ excitations and find that all the physical properties converge at $n=6$. Therefore, we 
demonstrate our results with $n=6$ (for comparison, in Ref.~\cite{Einax2011} the photons and phonons were treated in a non-self-consistent mean-field approximation). The 
expression for the current is obtained from the formal continuity equation $\frac{\d \hat{n}_{D,1}}{\d t}=\hat{J}_{\mathrm{in}}-\hat{J}_{\mathrm{out}}$, where $\hat{n}_{D,1}=c^{\dagger}_{D,1}c_{D,1}$. The resulting expression for the current is $J\equiv  \langle \hat{J}_{\mathrm{out}}\rangle=\langle \hat{J}_{\mathrm{in}}\rangle=2 \gamma_L \left( f_L(\e_{D,1})-\langle \hat{n}_{D,1} \rangle \right)$, where $\langle \cdot \rangle$ represents the steady state average.  Equivalently, the current $\langle \hat{J}_{\mathrm{out}}\rangle$ can be written as the sum of photon-induced and phonon-induced current, $J=2 \mathrm{Im}\langle ( \lambda_{e-pht}  a^{\dagger} +\lambda_{e-phn}  b^{\dagger}) c^\dagger_{D,1} c_{D,2} \rangle$.

\section{Acknowledgements} 
The authors thank I. Visoly-Fisher for valuable discussions. SA and YD acknowledge support from a BGU start-up grant. BZ acknowledges the FONDECYT grant no.  3130495.

 \section{Additional information}
 {\sl Competing financial interests}: the authors declare no competing financial interests. 

 \section{Author contributions}
 Y.D. conceived the research and drafted the paper. S.A. performed the numerical and analytical work together with B.Z. All authors 
 discussed the results and contributed to the final version of the manuscript.

\end{document}